\documentclass[pr,reprint,floatfix,superscriptaddress]{revtex4-1}
\usepackage{amsmath,amssymb}
\usepackage{float}
\usepackage{graphicx}

\usepackage{ulem}

\usepackage[english]{babel}
\usepackage[utf8]{inputenc}
\usepackage{hyperref}
\usepackage{color}
\hypersetup{colorlinks=true,citecolor={blue},linkcolor={blue},urlcolor={blue}}
\usepackage{natbib}

\usepackage{bm}

\hypersetup{colorlinks=true,citecolor={blue},linkcolor={blue},urlcolor={blue}}
\graphicspath{{images/}}

\renewcommand{\i}{\mathrm i}
\newcommand{\dd}[2]{\frac{\partial #1}{\partial #2}}

\begin{document}

\title{Inverse-phase Rabi oscillations in semiconductor microcavities}
\author{A.~V.~Trifonov}
\email[correspondence address: ]{arthur.trifonov@gmail.com}
\affiliation{Spin Optics Laboratory, St. Petersburg State University, St. Petersburg, 198504, Russia
}
\author{N.~E.~Kopteva}
\affiliation{Spin Optics Laboratory, St. Petersburg State University, St. Petersburg, 198504, Russia
}
\author{M.~V.~Durnev}
\affiliation{Ioffe Institute, St. Petersburg, 194021, Russia
}
\author{I.~Ya.~Gerlovin}
\affiliation{Spin Optics Laboratory, St. Petersburg State University, St. Petersburg, 198504, Russia
}
%\author{I.~V.~Ignatiev}
%\affiliation{Spin Optics Laboratory, St. Petersburg State University, St. Petersburg, 198504, Russia
%}
\author{R.~V.~Cherbunin}
\affiliation{Spin Optics Laboratory, St. Petersburg State University, St. Petersburg, 198504, Russia
}
\author{A.~Tzimis}
\affiliation{ Department of Materials Science and Technology, University of Crete,
71003 Heraklion, Crete, Greece
}
\affiliation{FORTH-IESL, PO Box 1385, 71110 Heraklion, Crete, Greece
}
\author{S.~I.~Tsintzos}
\affiliation{ Department of Materials Science and Technology, University of Crete,
71003 Heraklion, Crete, Greece
}
%\affiliation{FORTH-IESL, PO Box 1385, 71110 Heraklion, Crete, Greece
%}
\author{Z.~Hatzopoulos}
\affiliation{ Department of Materials Science and Technology, University of Crete,
71003 Heraklion, Crete, Greece
}
%\affiliation{FORTH-IESL, PO Box 1385, 71110 Heraklion, Crete, Greece
%}
\author{P.~G.~Savvidis}
\affiliation{ Department of Materials Science and Technology, University of Crete,
71003 Heraklion, Crete, Greece
}
\affiliation{FORTH-IESL, PO Box 1385, 71110 Heraklion, Crete, Greece
}
\author{A.~V.~Kavokin}
\affiliation{ School of Physics and Astronomy, University of Southampton, Southampton, SO17 1BJ, United Kingdom
}
\affiliation{Spin Optics Laboratory, St. Petersburg State University, St. Petersburg, 198504, Russia
}

\date{\today }

\begin{abstract}

We study experimentally the oscillations of a non stationary transient signal of a semiconductor microcavity  with embedded InGaAs quantum wells. The oscillations occur as a result of quantum beats between the upper and lower polariton modes due to the strong exciton-photon coupling in the microcavity sample (Rabi oscillations). The registration of spectrally resolved signal has allowed for separate observation of oscillations at the eigenfrequencies of two polariton modes. Surprisingly, the observed oscillations measured at the lower and upper polariton modes have opposite phases. We demonstrate theoretically that the opposite-phase oscillations are caused by the pump-induced modification of polariton Hopfield coefficients, which govern the ratio of exciton and photon components in each of the polartion modes. Such a behaviour is a fundamental feature of the quantum beats of coupled light-matter states. In contrast, the reference pump-probe experiment performed for the pure excitonic states in a quantum well heterostructure with no microcavity revealed the in-phase oscillations of the pump-probe signals measured at different excitonic levels.

\end{abstract}

\maketitle

%Abstract complete,insert pacs.

\section{Introduction}

Microcavity embedded semiconductor nanostructures have been subject of intense studies during recent years. Due to the high density of the photon field inside a microcavity these structures feature exceptionally strong light-matter coupling~\cite{Weisbuch}. The effects of strong coupling might be applied for realization of low-threshold lasers~\cite{Schneider:2013qr, PhysRevLett.110.206403}, logic elements for optical computers~\cite{Ballarini:2013}, memory elements for quantum computations~\cite{PhysRevLett.99.196402, Demirchyan}, sources of terahertz emission~\cite{PhysRevLett.110.047402}, etc. The light-matter coupling is the most efficient at the resonance between cavity mode and the exciton transitions inside the cavity. This resonance results in formation of coupled photon-exciton excitations -- exciton-polaritons.

The polariton effect might be observed in various excitonic systems~\cite{Razbirin}, however it is most pronounced in the microcavity structures. In high-finesse microcavities the energy of exciton-photon interaction is enhanced by about three order of magnitude as compared to bulk semiconductor materials and thin films. Exciton-photon coupling in microcavities leads to formation of lower (LP) and upper (UP) polariton modes, which are split by $\sim 10$~meV in typical GaAs  based samples. Excitation of both polariton modes with a short optical pulse leads to oscillations of exciton polarization and electric field amplitudes at the frequency defined by the value of the splitting between polariton modes (the so-called, vacuum Rabi oscillations)~\cite{Norris, PhysRevB.56.7564, PhysRevB.55.7084, PhysRevB.58.7978, MV_Rabi, PhysRevB.90.245309, PhysRevLett.112.113602, PhysRevLett.113.226401,Colas2015}.

The study of Rabi oscillations allows for a detailed understanding of polariton dynamics in microcavity structures~\cite{PhysRevLett.113.226401,Colas2015,PhysRevB.92.235305, PhysRevB.94.195301, PhysRevB.92.125415}. Since the typical values of Rabi oscillation period lie in a pico- or sub picosecond range, the most effective way to study the Rabi oscillations is to apply the methods of coherent spectroscopy with femtosecond laser pulses. These methods, generally called pump-probe, are based on probing with a probe pulse the changes in medium properties induced by a pump pulse, and have been used for a long time to study the exciton dynamics in semiconductor nanostructures~\cite{Shah1999}.

The experimental data obtained with the pump-probe technique is usually successfully described by means of optical Bloch equations, where the coupling between the pump and probe pulses is accounted for in the dielectric susceptibility of the structure by non-linear in the light amplitude terms~\cite{Shah1999}. The physical processes responsible for the nonlinearity in semiconductor nanostructures are depletion of the ground state due to the Pauli blockade and Coulomb screening of excitons, which reduce the oscillator strength of the exciton transition. The depletion of the ground state by pump field is a common effect for two-level quantum systems, whereas the exciton-exciton interaction is typical for semiconductors only.

The dynamics of polaritons in microcavities has a lot in common with dynamics of other excitonic systems, however, due to the coupled light-matter nature of the polariton states, there are substantial differences in their behaviour. In particular, in contrast to conventional nanostructures, where after excitation with a short light pulse the excitons evolve in the absence of the electromagnetic field, the electromagnetic field in microcavities is present during the whole lifetime of polaritons. As a result, additional effects, such as the blue shift of polariton energy levels under intense optical excitation, appear. As we will show in this work, it is mainly the blue shift effect, that is responsible for formation of Rabi oscillations in a pump-probe signal.

We have investigated the transient optical response of a semiconductor microcavity with InGaAs quantum wells (QWs) using spectrally resolved pump-probe technique. We have revealed distinct Rabi oscillations of the optical signals at each of the polariton modes. For all excitation conditions we tried the observed oscillations corresponding to lower and upper polariton modes are found to have opposite phases. In order to identify the nonlinear effects that contribute to the formation of the measured signal we have developed a theoretical model.

%Статья организована следующим образом. В следующем разделе представлены детали эксперимента и структура исследованного образца. Затем представлены экспериментальные данные. После этого следует теоретическое описание наблюдаемых в эксперименте зависимостей и обсуждение полученных результатов. В обсуждении приводится сравнение механизмов формирования осцилляций нестационарного сигнала квантовых биений в обычных экситонных структурах и поляритонных осцилляций Раби в структурах с микрорезонатором.

\section{Experiment}

The experimentally studied sample is a relatively low-finesse microcavity ($Q\approx$~2000), which consists of 17 and 21 pairs of AlAs/GaAs layers. The four groups of pairs of In$_x$Ga$_{1-x}$As quantum wells with different concentration of In ($x = 0.08$ and $x = 0.12$) are embedded inside the microcavity. The cavity length is variable across the sample plane, which allows one to change the detuning between the exciton and photon modes. The photon mode is tuned to the resonance with a deeper well with $x = 0.12$. The distance between the neighboring QWs is sufficiently large, so that one can neglect the effects of the inter-well carrier tunneling and coupling. The substrate is transparent in the spectral range of QW excitonic resonance, which allows one to study the sample in the transmission geometry. 

\begin{figure}
	\centering
	%\begin{subfigure}[b]{1\linewidth}
        		\includegraphics[width=\linewidth]{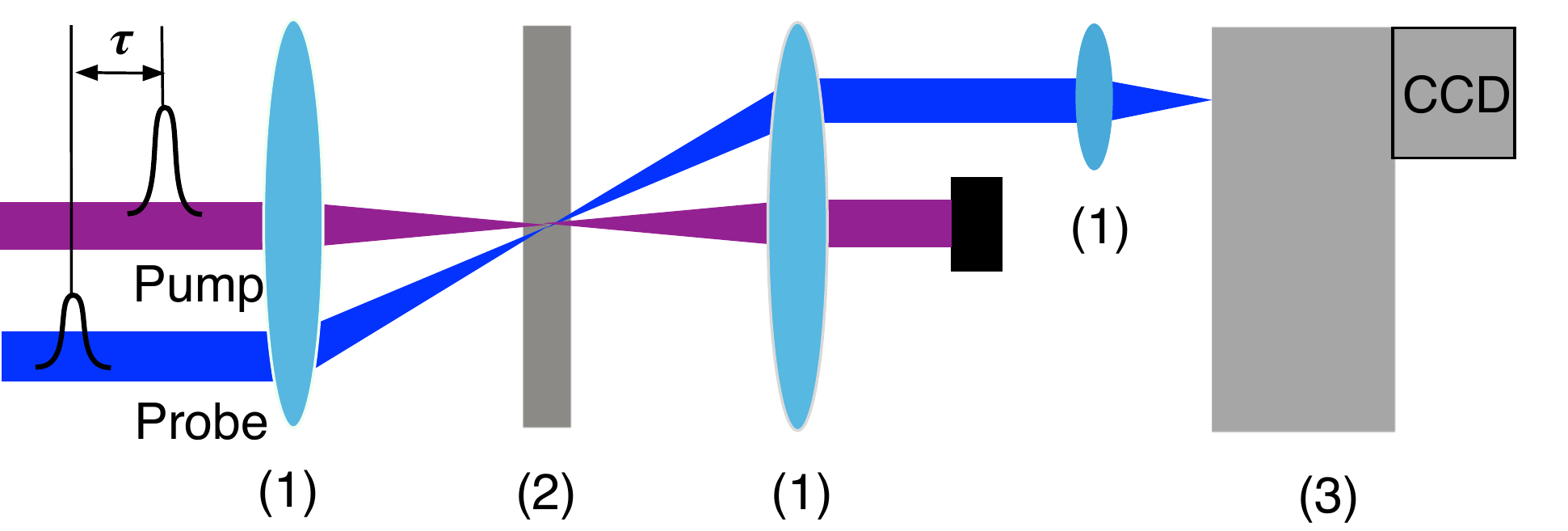}
        %\end{subfigure}
	\caption{\label{fig1} Schematic sketch of the experimental setup that comprises the lenses (1), the sample (2) and the spectrometer with a CCD-camera (3). The probe beam is incident at a 5$^\circ$ angle with respect to the sample normal. The probe pulses are delayed by the time $\tau$ with respect to the pump pulse. 
	} 
\end{figure}

The sample was cooled down to the temperature of 5~K in a closed-cycle helium cryostat. In the experimental setup, see Fig.~\ref{fig1}, the emission of femtosecond titan-sapphire laser with the duration of 100 fs and the repetition frequency of 80 MHz was split into two beams. The first, pump, beam was directed along the sample normal and focused with a lens with a focal length of 150 mm into the spot with a 70~$\mu$m diameter. The second, probe, beam was incident at a 5$^\circ$ angle with respect to the sample normal, passing through the delay line and focused into a spot of 30~$\mu$m diameter.

The probe emission passing through the sample was focused into spectrometer and detected by a nitrogen-cooled  CCD-camera. The CCD-camera scan was synchronized with a motion of the delay line that allowed us to observe the transmission spectrum as a function of the delay time between the probe and pump pulses.

%\section{Экспериментальные результаты}

\begin{figure}[htbp]
\centering
\includegraphics[width=0.97\linewidth]{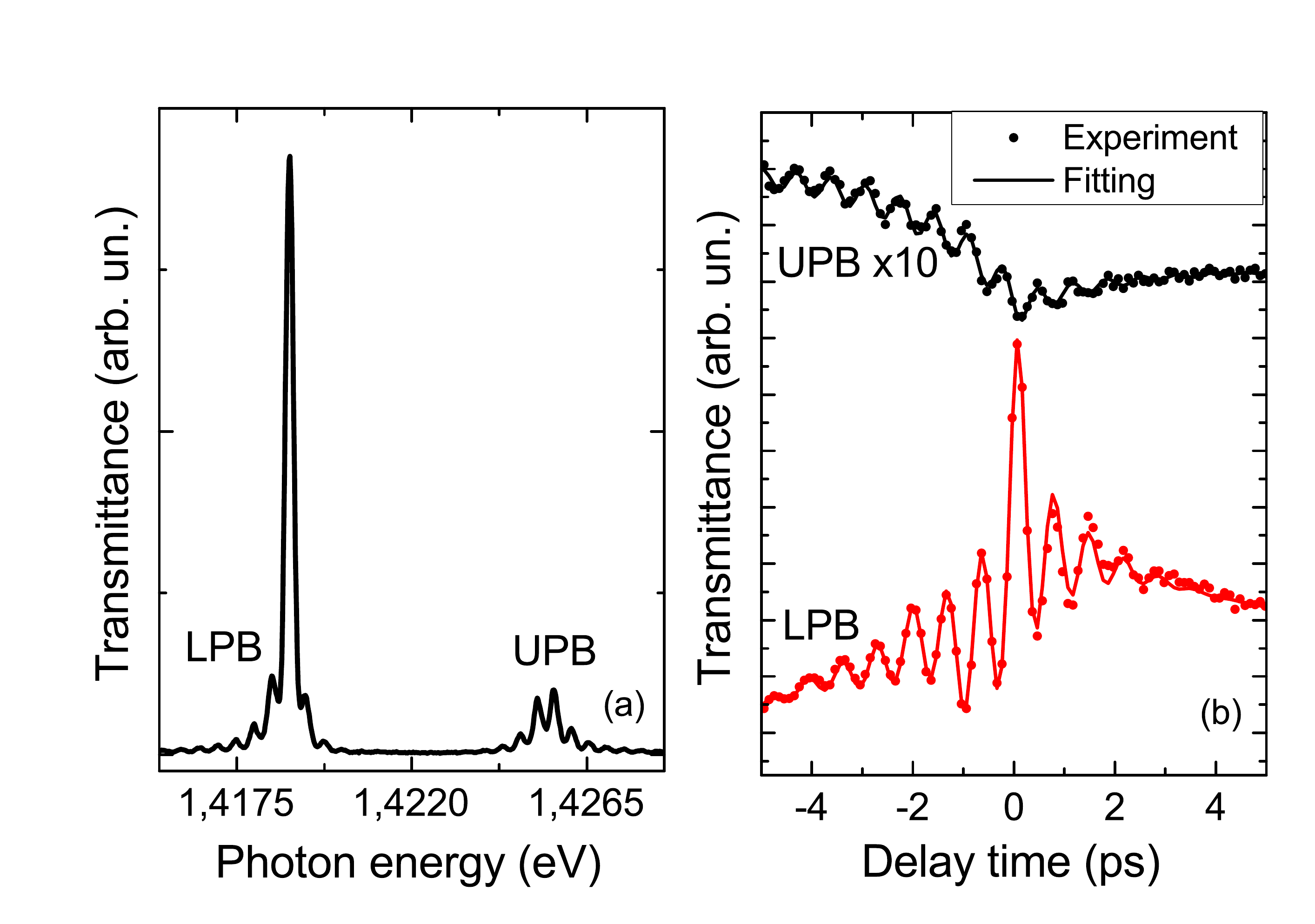}
\caption{ (a)  Transmission spectrum of the studied sample. (b) Intensities of the spectral peaks in the probe transmission spectrum as a function of the delay time between the pump and probe pulses. The black and red dots show the data for the UP and LP frequencies, respectively, the solid lines are fits (see the text for details). Data for the UP mode is multiplied by a factor of 10. The data is measured at minimal intensities of the pump and probe beams, used in the experiment ($P_{pu}=0.35$~mW and $P_{pr}=0.1$~mW), which correspond to the polariton density $\lesssim 10^9$~cm$^{-2}$.}
\label{fig2}
\end{figure}

In the transmission spectrum of the sample, see Fig.~\ref{fig2}~(a), two narrow peaks are observed in the range of anti crossing of a polariton dispersion. These two peaks correspond to the upper and lower polariton modes, which are formed in the sample. The investigations were carried out at the sample point with a negative detuning, $\Delta = -3$~meV, between the cavity and exciton modes. The short-period oscillations observed at the spectral contours are related to the light interference in the sample substrate. With increase of the excitation power the LP peak shifts to higher energies, indicating the effects of polariton-polariton interaction. 
 
% На рисунке~\ref{fig1}~(b) черными точками представлена зависимость величины голубого сдвига от мощности накачки, измеренная при нулевой задержке между co-циркулярно поляризованными импульсами накачки и зондирования. Сплошной кривой показана аппроксимация функцией .... Как будет показано в обсуждении, из этих данных можно извлечь величину констант поляритон-поляритонного взаимодействия \textit{Правильно ли я понимаю, что из зависимости голубого сдвига от мощности накачки можно извлечь абсолютную величину параметра взаимодействия поляритонов?}.

The dots in Fig.~\ref{fig2}~(b) show the dependence of the signal intensity, detected at UP (black dots) and LP (red dots) modes, as a function of the delay time $\tau$ between pump and probe pulses. At positive delay time, $\tau > 0$, the pump pulse arrives before the probe one, while $\tau < 0$ corresponds to the first arrival of the probe pulse.
% При этом детектировалось только излучение в направлении прошедшего через образец луча зондирования. 
The experimental data was approximated with the following phenomelogical functions:
 \begin{multline}
 \label{eq:fit}
f_{u,l}^{\pm}(\tau) = A_{u,l}^{\pm} + B_{u,l}^{\pm} \exp(\mp \tau/t_{u,l}^{\pm})  +\\
+ C_{u,l}^{\pm} \cos(\Omega_{u,l}^{\pm}\tau+\varphi_{u,l}^{\pm})\exp(\mp \tau/T_{u,l}^{\pm})\:,
\end{multline}
where the subscripts $u$ and $l$ denote the upper and lower polariton modes, and the superscripts $+$ and $-$ denote the regions of positive and negative $\tau$, respectively.
The presented curves are measured at a relatively small pump power, which corresponds to the polariton density $\lesssim 10^9$~cm$^{-2}$. The blue shift of spectral lines in this case is not resolved. 

\begin{figure}[htbp]
\centering
\includegraphics[width=1\linewidth]{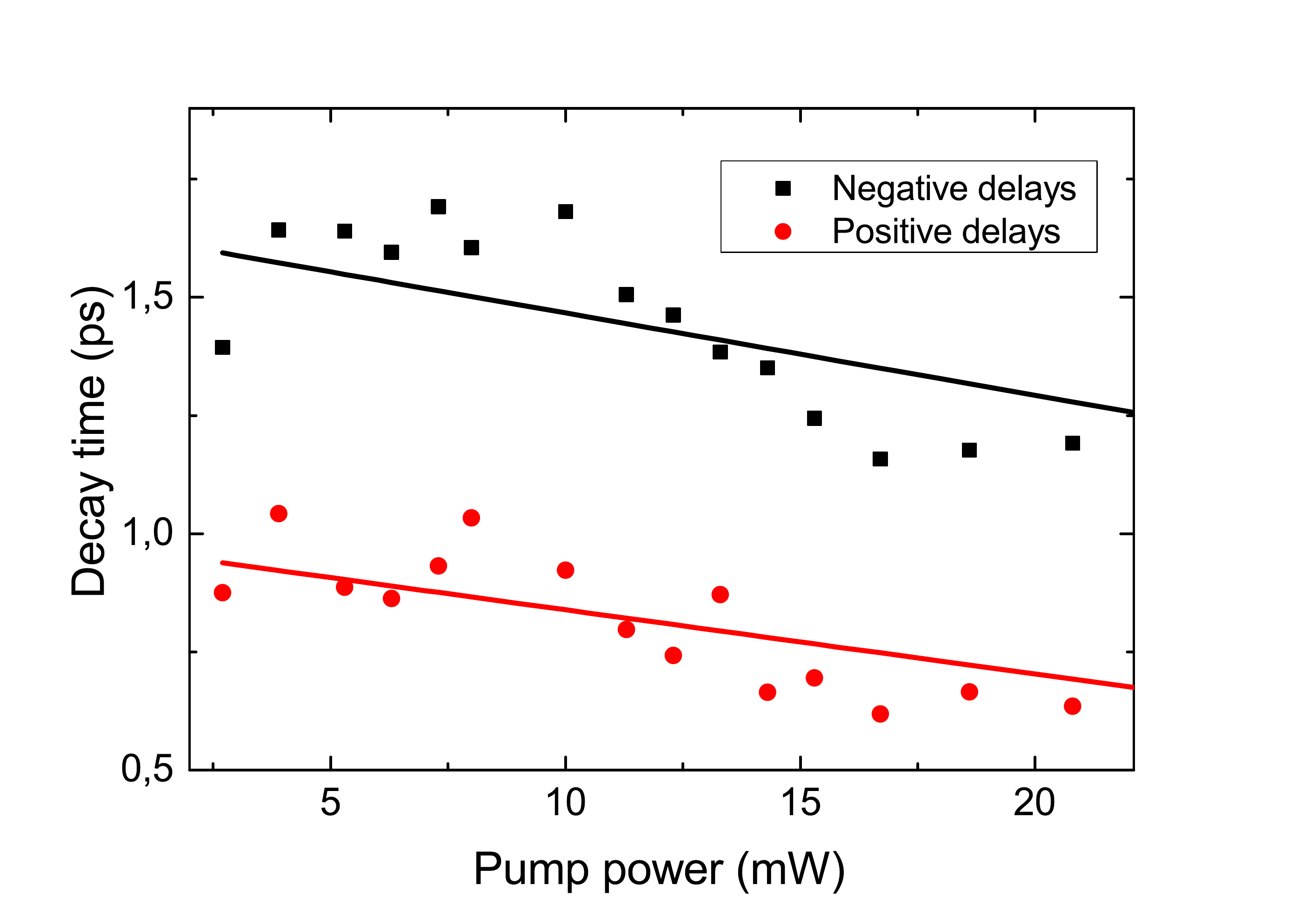}
\caption{Oscillations decay time as a function of pump power at the lower polariton branch. The solid lines are guides for the eye.
}
\label{fig3}
\end{figure}

As seen from Fig.~\ref{fig2}, the oscillations of signal are observed at positive as well as at negative delay time. Moreover, the oscillations at $\tau <0$ decay slower than at $\tau >0$, which becomes most pronounced at high pump powers, see Fig.~\ref{fig3}. Such an asymmetry of the signal decay is typical for nonlinear optical phenomena driven by the third-order nonlinearities~\cite{ivchenko05a}.

The main result of this work is that the Rabi oscillations observed at lower and upper polariton modes have an opposite phase, see Fig.~\ref{fig2}. This result is valid for all values of pump and probe power used in our experiment. To understand the origin of the inverse-phase oscillations and to specify the nature of the signal at negative delays we performed the theoretical analysis of polariton dynamics in pump-probe experiments. The results of this analysis are presented in the next section.

\section{Theory}

To analyze the dynamics of polaritons in our sample we use coupled nonlinear equations which describe evolution of the photon mode in the microcavity and the exciton polarization in QWs~\cite{PhysRevB.55.9872, PhysRevB.92.235305, PhysRevB.61.13856, carmichael_book}:
\begin{eqnarray}
\label{eq:system}
\dd{P}{t} &=& \left[ -\i ( \omega_x - \bar{\omega} + \alpha N_x ) - \Gamma \right] P -\i (\Omega_R -\beta N_x) E\:, \nonumber \\
\dd{E}{t} &=& \left[ -\i (\omega_c - \bar{\omega}) - \gamma \right] E -\i \Omega_R P + \mathcal E(t)\:.
\end{eqnarray}
Here $E(t)$ is the slowly-varying amplitude of the electric field in the center of the QW, $P(t)$ is the slowly-varying amplitude of the excitonic polarization averaged over the QW width, $\omega_x$ and $\omega_c$, and $\Gamma$ and $\gamma$ are the resonance frequencies and decay rates of exciton and photon modes, respectively, $\Omega_R$ is the Rabi frequency, which determines the coupling of exciton and photon modes, $\mathcal E(t)$ and $\bar{\omega}$ are the amplitude and the optical frequency of the incident electric field, $N_x = |P|^2$ is the exciton population, and $\alpha$ and $\beta$ are real parameters. Equations~\eqref{eq:system} are analogous to optical Bloch equations that describe the four-wave mixing in bulk materials~\footnote{We note that to describe the evolution of the excitonic mode the so-called excitonic Bloch equations were used in recent works~\cite{PhysRevB.61.13856, PhysRevB.92.235305}. In this approach there is a third equation, which accounts for additional decay channels of exciton population $N_x$ as compared to polarization decay $\Gamma$. The inclusion of these addition channels does not change general conclusions of our work, therefore we neglect them and use $N_x = |P|^2$.}.

The incident electric field is a sum of pump and probe fields inside a cavity, $\mathcal E = \mathcal E_1 \exp(\i \bm k_1 \bm r) + \mathcal E_2 \exp(\i \bm k_2 \bm r)$, where $\bm k_1$ and $\bm k_2$ are the wave vectors of pump and probe fields inside the cavity, $\bm r$ is a coordinate. We model the amplitudes $\mathcal E_{1,2}$ to be proportional to delta-functions in time domain: $\mathcal E_1 = \mathcal N_1 \delta(t - t_1)$, $\mathcal E_2 = \mathcal N_2 \delta(t - t_2)$, where $\mathcal N_1$ and $\mathcal N_2$ are the amplitudes, which we assume to be real, and $t_2 - t_1 = \tau$ is the delay time.

Nonlinear terms proportional to $N_x$ enter the first equation in Eqs.~\eqref{eq:system} and describe two possible nonlinearities in our system. The first one $\propto \alpha |P|^2 P$ is the blue shift of the exciton mode (or the so called anharmonic-like nonlinearity), and the second one $\propto \beta |P|^2 E$ is the reduction of the Rabi frequency (or the so-called two-level-like nonlinearity)~\cite{PhysRevB.55.9872, PhysRevA.44.2124}. These nonlinear terms 
%given by
%\begin{equation}
%F_{nl} = 2(g_0 P - g_1 \Omega_R  E) |P|^2
%\end{equation}
 result in the coupling of pump and probe signals, and consequently in Rabi oscillations of the output field. 
 
The electric field and polarization inside the cavity are the sum of the pump and probe components, $E = E_1 \exp(\i \bm{k}_1 \bm r) + E_2 \exp (\i \bm k_2 \bm r)$, and $P = P_1 \exp(\i \bm{k}_1 \bm r) + P_2 \exp (\i \bm k_2 \bm r)$. In the limit of small nonlinearities, $\alpha N_x, \beta N_x \ll \Omega_R$, approximate solutions of Eqs.~\eqref{eq:system} have a form 
\begin{equation}
E_{1,2} = \bar{E}_{1,2} + \delta E_{1,2}\:,\:\:\: P_{1,2} = \bar{P}_{1,2} + \delta P_{1,2}\:,
\end{equation}
where $\bar{E}_{1,2}$ and $\bar{P}_{1,2}$ satisfy Eqs.~\eqref{eq:system} at $\alpha = \beta = 0$, and $\delta P_{1,2}$ and $\delta E_{1,2}$ are small corrections due to the presence of nonlinearities. 

The time evolution of $\bar{E}_{1,2}$  and $\bar{P}_{1,2}$ is given by~\cite{PhysRevB.55.9872}
\begin{eqnarray}
\label{eq:PE}
\bar{P}_{j}(t) &=& -\i\mathcal N_{j} \frac{\Omega_R}{\tilde{\Omega}_R} \sin \tilde{\Omega}_R(t -t_j) \\
&\times&\exp\left[- \frac{\tilde{\gamma} - \i \Delta}{2} (t - t_j)\right] \theta (t - t_j)\:, \nonumber \\
\bar{E}_{j}(t) &=& \mathcal N_{j} \left[ \cos \tilde{\Omega}_R(t -t_j) -\i \frac{\Delta}{2 \tilde{\Omega}_R} \sin \tilde{\Omega}_R (t -t_j) \right] \nonumber \\
&\times&\exp\left[- \frac{\tilde{\gamma} - \i \Delta}{2} (t - t_j)\right] \theta (t - t_j)\:. \nonumber
\end{eqnarray}
Here $\tilde{\Omega}_R = \sqrt{\Omega_R^2 + \Delta^2/4}$ and $\Delta = \omega_c - \omega_x$ is the cavity mode detuning, $\tilde{\gamma} = (\gamma + \Gamma)/2$, $\theta (t)$ is a Heaviside function, and we assume $\bar{\omega} = \omega_c$. In the derivation of Eqs.~\eqref{eq:PE} we neglected small terms proportional to $\gamma/\Omega_R \ll 1$ and $\Gamma/\Omega_R \ll 1$.

The nonlinear terms proportional to $\exp(\i \bm k_2 \bm r)$, which give rise to Rabi oscillations in the direction of probe pulse, are 
\begin{equation}
\label{eq:nonlin}
F_{nl} = -\i \alpha |{P}_1|^2 {P}_2 + \i \beta \left( |{P}_1|^2 {E}_2 + {P}^*_1 P_2 E_1 \right)\:.
\end{equation} 
In the experiment we measure the Fourier transform of the output signal in the direction of the probe pulse, which is proportional to
\begin{equation}
\label{eq:Iw}
I_\omega = |E_{2, \omega}|^2 = \left| \bar{E}_{2,\omega} \right|^2 + \bar{E}_{2,\omega} \left( \delta E_{2,\omega} \right)^* + \left( \bar{E}_{2,\omega} \right)^* \delta E_{2,\omega}\:,
\end{equation}
where $F_\omega$ is the Fourier transform of $F$, $F_{ \omega} = \int F \exp(\i \omega t) dt$, and the star sign means the complex conjugate. In Eq.~\eqref{eq:Iw} we neglected a small contribution $|\delta E_{2,\omega}|^2$, which is proportional to the second power of nonlinearities $\alpha$ and $\beta$. Note that the first term in Eq.~\eqref{eq:Iw} is a transmission coefficient of the microcavity structure and does not depend on $\tau$.

Solving Eqs.~\eqref{eq:system} in the frequency space we find for the Fourier transforms $\bar{E}_{2,w}$ and $\delta E_{2,\omega}$:
\begin{eqnarray}
\label{eq:E2w}
\bar{E}_{2,w} &=& \i \mathcal N_2 \frac{\omega + \Delta + \i \Gamma}{(\omega + \Delta + \i \Gamma) (\omega + \i \gamma) - \Omega_R^2}\:, \\
\delta E_{2,\omega} &=& \frac{\i \Omega_R F_{nl, \omega}}{( \omega + \Delta + \i\Gamma) ( \omega + \i \gamma) - \Omega_R^2} \nonumber\:,
\end{eqnarray}
where $F_{nl, \omega}$ is a Fourier transform of Eq.~\eqref{eq:nonlin}.
Using Eqs.~\eqref{eq:Iw},~\eqref{eq:E2w} we find for the oscillating transient signal at upper and lower polariton modes
\begin{equation}
\label{eq:Iul}
I_{u,l} \equiv I_\omega (\omega_{u,l}) = \frac{2 \mathcal N_2 \Omega_R \left( \omega_{u,l} + \Delta \right)}{\left( 2\tilde{\gamma} \omega_{u,l} + \gamma \Delta\right)^2} \mathrm{Re} \left\{F_{nl, \omega} \left( \omega_{u,l} \right)\right\}\:,
\end{equation}
where $\omega_{u,l} \approx -\Delta/2 \pm \tilde{\Omega}_R$ are real parts of upper and lower polariton modes under the conditions $\gamma/\Omega_R \ll 1$, $\Gamma/\Omega_R \ll 1$, which are fulfilled in the experiment, and $\mathrm{Re}$ denotes the real part of a complex number. 

It follows from Eqs.~\eqref{eq:PE}, \eqref{eq:nonlin}, \eqref{eq:Iul} that the intensities $I^+_{u,l}$ at $\tau > 0$ and $I^-_{u,l}$ at $\tau < 0$ have a form
\begin{multline}
\label{eq:IRabi}
I_{u,l}^\pm = \frac{I_0  \mathrm{e}^{-|\tau|/T^\pm} \left( \omega_{u,l} +  \Delta \right)}{\left( 2\tilde{\gamma} \omega_{u,l} + \gamma \Delta\right)^2}  \times \\
 \left[ A^\pm_{u,l} + B^\pm_{u,l} \sin 2 \tilde{\Omega}_R \tau + C^\pm_{u,l} \cos 2 \tilde{\Omega}_R \tau \right]\:,
\end{multline}
%\begin{eqnarray}
%I_{u,l}^+ &=&\frac{I_0  \mathrm{e}^{-\tilde{\gamma} \tau}}{\omega_{u,l} + \Delta} \left[ A^+_{u,l} + B^+_{u,l} \sin 2 \tilde{\Omega}_R \tau + C^+_{u,l} \cos 2 \tilde{\Omega}_R \tau \right]\:, \nonumber\\
%I_{u,l}^- &=& \mathrm{e}^{\gamma \tau/2} \left[ A^-_{u,l} + B^-_{u,l} \sin 2 \tilde{\Omega}_R \tau + C^-_{u,l} \cos 2 \tilde{\Omega}_R \tau \right] \:.
%\end{eqnarray}
%Here superscripts $\pm$ denote the signals at $\tau > 0$ and $\tau <0$, respectively, 
where $I_0 = -2 \mathcal N_1^2 \mathcal N_2^2 \Omega_R^4/\tilde{\Omega}_R^4$.
 The calculations yield 
 \begin{equation}
 \label{eq:decay}
T^+ = 1/\tilde{\gamma}\:,\:\:\: T^- = 2/\tilde{\gamma}\:, 
\end{equation}
\begin{eqnarray}
\label{eq:coeffs}
A_{u,l}^+ = &\dfrac{1}{8} \left( \alpha + \dfrac{\pm \tilde{\Omega}_R - \Delta}{\Omega_R} \beta \right) \:,\:\:\: B_{u,l}^+ = \dfrac{\tilde{\Omega}_R}{12 \tilde{\gamma}} \left( \alpha -\dfrac{\Delta}{\Omega_R} \beta \right) \:,\nonumber  \\
 &C_{u,l}^+ = -\dfrac{1}{32} \left( \alpha - \dfrac{\pm2\tilde{\Omega}_R + \Delta}{\Omega_R} \beta \right)\:.
\end{eqnarray}
and
\begin{eqnarray}
\label{eq:coeffsmin}
A_{u,l}^- = &\pm \dfrac{\beta}{8} \:,\:\:\: B_{u,l}^- = \dfrac{\tilde{\Omega}_R}{12 \tilde{\gamma}} \left( \alpha -\dfrac{\Delta}{\Omega_R} \beta \right) \:,\nonumber  \\
 &C_{u,l}^- = \dfrac{1}{32} \left( 3\alpha + \dfrac{\pm2\tilde{\Omega}_R - 3\Delta}{\Omega_R} \beta \right)\:.
\end{eqnarray}

\begin{figure}[htbp]
\centering
\includegraphics[width=0.49\linewidth]{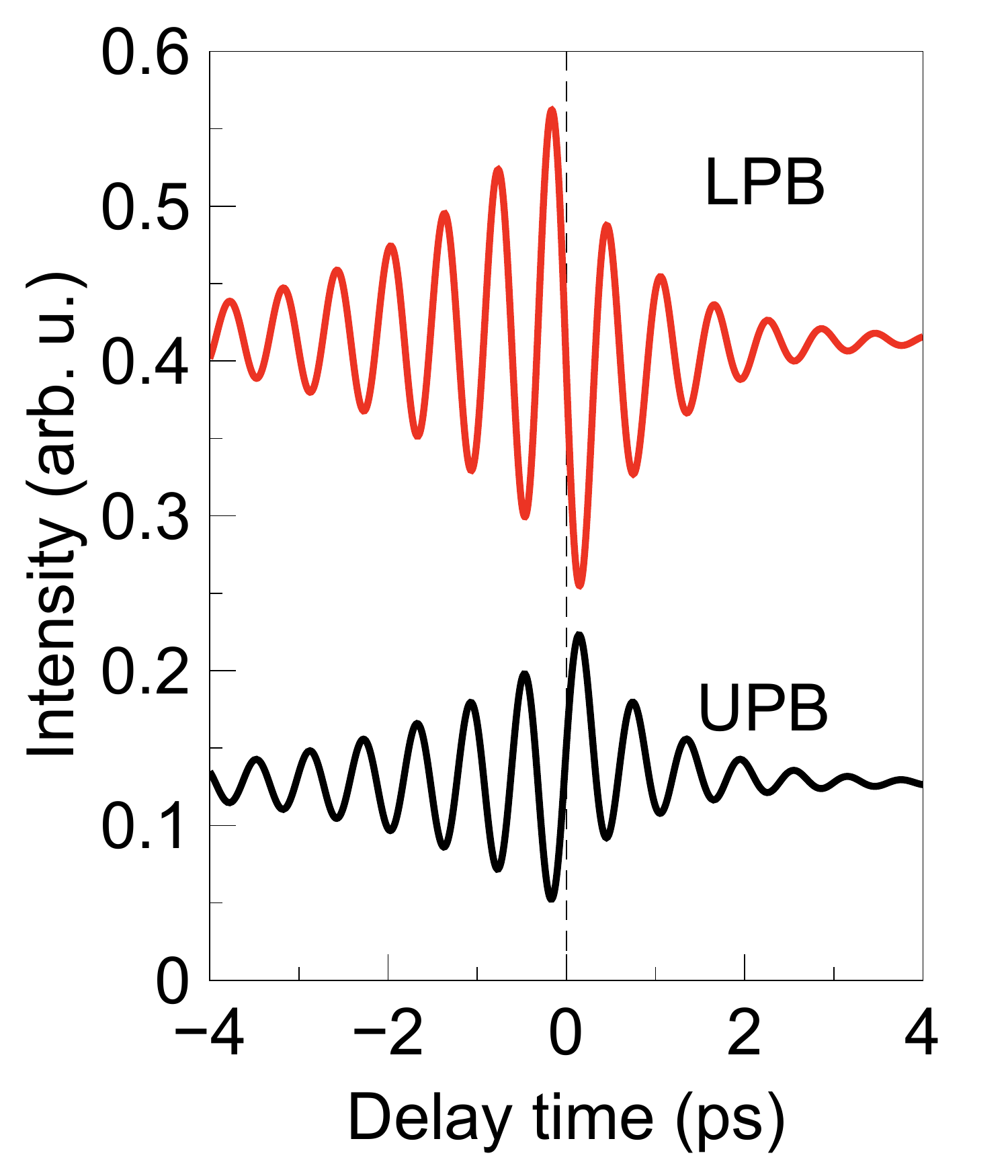}
\includegraphics[width=0.49\linewidth]{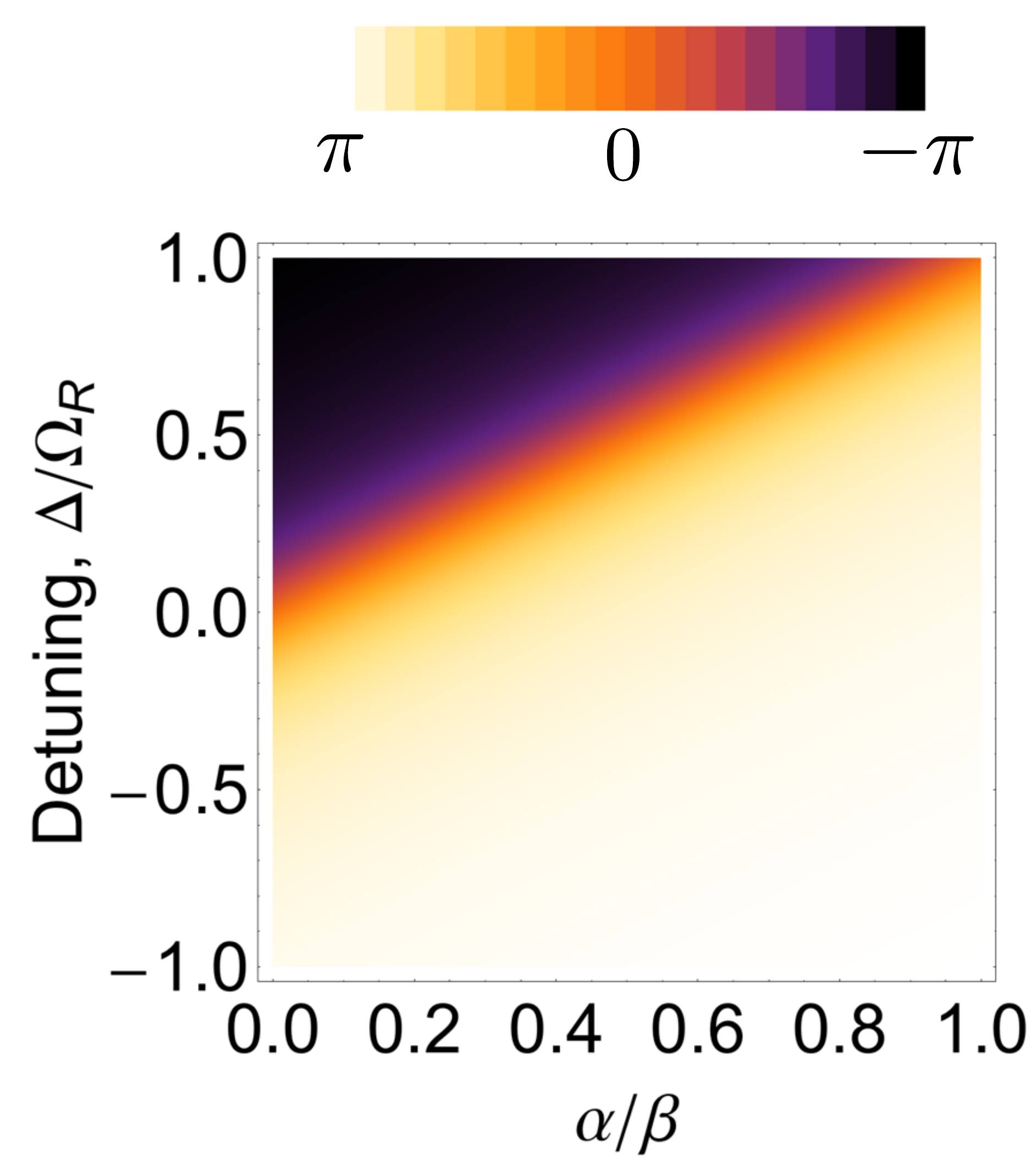}
\caption{(Left panel) Transient signals calculated using Eqs.~\eqref{eq:IRabi}, \eqref{eq:coeffs}, \eqref{eq:coeffsmin}. The parameters used are $\Omega_R = 5$~ps$^{-1}$, $\Delta = -3$~ps$^{-1}$, $\gamma = \Gamma = 1$~ps$^{-1}$, and $\alpha/\beta = 5$. (Right panel) Phase shift between Rabi oscillations at LP and UP modes at $\tau > 0$. $\Omega_R = 5$~ps$^{-1}$, $\gamma = \Gamma = 1$~ps$^{-1}$.
}
\label{fig4}
\end{figure}

With the use of Eqs.~\eqref{eq:IRabi}, \eqref{eq:coeffs}, \eqref{eq:coeffsmin} let us now analyze the behaviour of Rabi oscillations for two types of nonlinearities given by Eq.~\eqref{eq:nonlin}, see Fig.~\ref{fig4}. For the anharmonic-like nonlinearity ($\beta = 0$, $\alpha \neq 0$) the intensities at the upper and lower modes $I_{u,l}^\pm$ differ only by the sign of the numerator in Eq.~\eqref{eq:IRabi}, which results in the opposite-phase oscillations. 

In the case of the two-level-like nonlinearity ($\alpha = 0$, $\beta \neq 0$) the situation is more complicated. It is seen from Eq.~\eqref{eq:coeffs} that for $\Delta \neq 0$ the main contribution to $I_{u,l}^\pm$ is given by $B_{u,l}^\pm$, which is parametrically large ($\tilde{\Omega}_R/\tilde{\gamma} \gg 1$). This results again in the opposite-phase oscillations. However if $\Delta = 0$ coefficient $B_{u,l}^\pm$ vanishes and the oscillations are governed by $C_{u,l}^\pm$, which gives the in-phase oscillations. 

In general the in-phase oscillations occur in the region of parameters when $B_{u,l}^\pm \ll C_{u,l}^\pm$, i.e. for $\alpha/\beta -\Delta/\Omega_R \ll \tilde{\gamma}/\tilde{\Omega}_R$. Since $\tilde{\gamma}/\tilde{\Omega}_R \ll 1$, this is a very narrow region in the vicinity of $\alpha/\beta = \Delta/\Omega_R$. In the remaining region of parameters the phase between oscillations on upper and lower modes equals to $\pm \pi$. We note that in our sample $\Delta < 0$ and therefore, since $\alpha, \beta >0$, the inverse-phase oscillations occur for any ratio between $\alpha$ and $\beta$.

\section{Discussion}

As a result of the strong light-matter coupling the excited states of a semiconductor quantum well embedded in a microcavity are superpositions of matter excitations (excitons) and electromagnetic field (photon mode). The intense light radiation not only affects the population of the ground and excited states of the system, but also changes the internal structure of these states. As seen from the theoretical analysis given above, it is the change of the internal structure of excitations that is the main source of nonlinearities, which result in the formation of the studied signals.

Each type of  nonlinearity in Eq.~\eqref{eq:nonlin} is accompanied by the change in Hopfield coefficients $C_{u,l}$~\cite{PhysRev.112.1555}, which define the photon and exciton contributions at upper and lower polariton states, and therefore the spectrum of the microcavity transmission coefficient.
For the microcavity polaritons, $C_{u,l}$ are controlled by the ratio of the exciton-photon detuning $\Delta$ and the Rabi splitting $\Omega_R$. The blue shift of exciton level, proportional to the interaction constant $\alpha$, alters the value of detuning $\Delta$, whereas the saturation of the oscillator strength, described by the parameter $\beta$, results in the reduction of $\Omega_R$. In the limit of small nonlinearities, $\alpha N_x, \beta N_x \ll \Omega_R$ we have
\begin{eqnarray}
|C_{u}|^2 &=& |C_u (0)|^2 - \frac{2(\alpha \Omega_R - \beta \Delta) \Omega_R}{(\Delta^2 + 4\Omega_R^2)^{3/2}} N_x\:, \nonumber \\
|C_{l}|^2 &=& |C_l (0)|^2 + \frac{2(\alpha \Omega_R - \beta \Delta) \Omega_R}{(\Delta^2 + 4\Omega_R^2)^{3/2}} N_x\:,
\end{eqnarray}
where $C_{u,l}(0)$ are the Hopfield coefficients at $N_x = 0$. Due to the property $|C_{u}|^2 + |C_{l}|^2 = 1$, any change in one coefficient results in the opposite change of another one, leading to opposite-phase Rabi oscillations of the pump-probe signal. 
However at a special condition $\alpha \Omega_R - \beta \Delta \approx 0$ the reduction of detuning with increase of $N_x$ is compensated by the reduction of $\Omega_R$, and $C_{u,l}$ are left unchanged. In this case the Rabi oscillations are matched in phase. This prediction is in agreement with the accurate theory of pump-probe signal, see Eqs.~\eqref{eq:coeffs}, \eqref{eq:coeffsmin}.

\begin{figure}
\centering
\includegraphics[width = 0.8\linewidth]{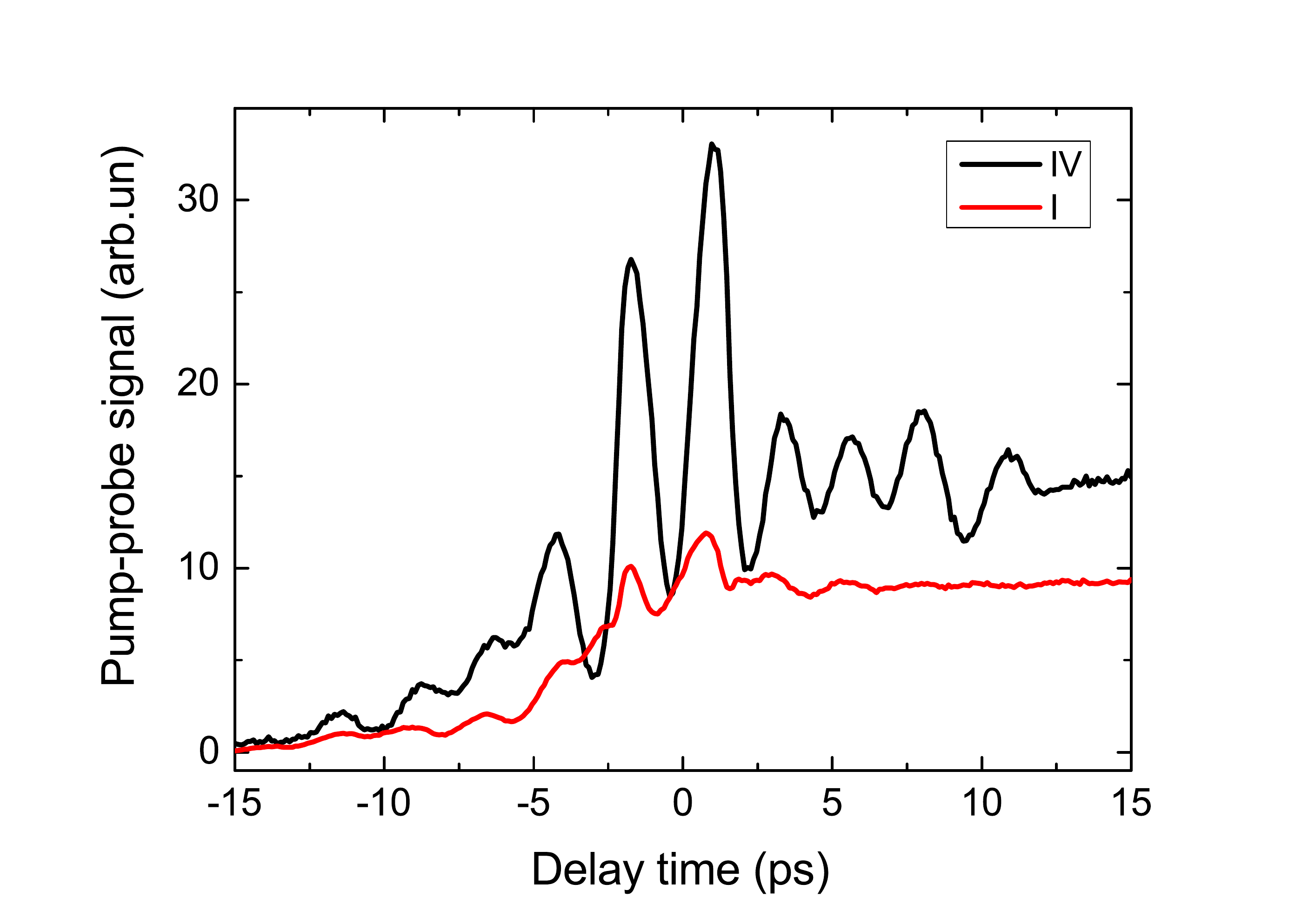}
\caption{Pump-probe signals of quantum beats between the I and IV quantum confined excitonic states in the 95~nm InGaAs/GaAs quantum well.  The signals were measured by tuning the detected photon energy at the I and IV excitonic resonances. The frequency of signal oscillations corresponds to the energy splitting between these states.}

\label{fig5}
\end{figure}

The opposite-phase signal in the region of positive delays (the standard pump-probe signal) is governed by the nonlinearity of the type $(P_1 P_1^*) P_2$, corresponding to the scattering of the probe pulse on the excitonic polarization created by the pump pulse. In the region of negative delays, the oscillating signal (the four-wave mixing signal) is governed by the nonlinearity $(P_2 P_1^*) P_1$, which has the same mathematical form, however in that case corresponds to the scattering of the pump pulse on the polarization pattern created by simultaneous action of pump and probe pulses. Such an asymmetry should result in the twofold increase of the signal decay time at $\tau < 0$ ($T^+$) compared to $\tau > 0$ ($T^-$), as confirmed by Eq.~\eqref{eq:decay}. Such a behaviour is observed in experiment in a wide range of pump powers for LPB and at low pump powers for UPB, see Fig.~\ref{fig3}. The increase of the $T^-/T^+$ ratio at higher pump powers for UPB might be attributed to the exciton-exciton scattering, which drives the phase relaxation of polarization.  

To confirm that the inverse-phase behaviour of Rabi oscillations is a specific feature of the microcavity systems with strong light-matter coupling we performed the same experiment for a quantum well structure without Bragg mirrors. We investigated the sample with an InGaAs quantum well of 95~nm width and 2~\% indium concentration.
In the absorption and photoluminescence spectra we observed series of narrow lines corresponding to optical transitions to or from the quantum-confined  excitonic levels. Detailed characteristics of the sample and experimental details can be found in Refs.~\cite{PhysRevB.91.115307, PhysRevB.92.201301}.

Figure~\ref{fig5} shows intensities of the pump-probe signals measured at energies corresponding to transitions to the first and fourth quantum-confined levels. The energy difference between the levels is approximately 2~meV, which is comparable to the Rabi splitting in the microcavity sample. One can see distinct signal oscillations, caused by quantum beats between the excitonic states. In contrast to the microcavity sample, the oscillation phases at two energies coincide, which is typical for quantum beats between matter excitations, such as excitons~\cite{PhysRevB.60.5797}. Indeed, for the excitonic quantum beats the microscopic mechanism of the pump-probe signal is related mainly to the depopulation of the ground state due to the pump beam excitation~\cite{PhysRevA.35.1720}. In this case the quantum beats observed  at optical transitions into two excited states have equal phases.

\section{Conclusion}

To conclude, we studied the transient pump-probe signal through the microcavity heterostructure with embedded InGaAs quantum well. The studies revealed well pronounced oscillations related to the quantum beats between lower and upper polariton modes (vacuum Rabi oscillations). The principle result is that the observed oscillations measured at lower and upper polariton levels have opposite phase. As revealed by the theoretical analysis, the opposite phases of oscillations are related to the specific light-matter character of polariton states in microcavity, in contrast to quantum beats between pure matter excitations. The experiments showed that, in agreement with theoretical prediction, the oscillations in the region of positive delay time between pump and probe pulses decay twice faster than in the region of negative delay time.

\begin{acknowledgments}

Financial support from the SPbGU (Grant No. 11.38.277.2014), the RFBR, and DFG in the frame of International Collaborative Research Center TRR 160 (Project No. 15-52-12018) is acknowledged. A. V. T. and R. V. C. are thankful the RFBR for the financial support in the frame of grant No. 15-59-30406-PT. M.V.D. was partially supported by RFBR project No. 16-32-60175, the Russian Federation President Grant No. MK-7389.2016.2 and the Dynasty foundation. The authors thank the SPbU Resource Center “Nanophotonics” (photon.spbu.ru) for providing the sample with wide quantum well studied in this work.

\end{acknowledgments}

\normalem

%\bibliography{literature}

\begin{thebibliography}{30}%
\makeatletter
\providecommand \@ifxundefined [1]{%
 \@ifx{#1\undefined}
}%
\providecommand \@ifnum [1]{%
 \ifnum #1\expandafter \@firstoftwo
 \else \expandafter \@secondoftwo
 \fi
}%
\providecommand \@ifx [1]{%
 \ifx #1\expandafter \@firstoftwo
 \else \expandafter \@secondoftwo
 \fi
}%
\providecommand \natexlab [1]{#1}%
\providecommand \enquote  [1]{``#1''}%
\providecommand \bibnamefont  [1]{#1}%
\providecommand \bibfnamefont [1]{#1}%
\providecommand \citenamefont [1]{#1}%
\providecommand \href@noop [0]{\@secondoftwo}%
\providecommand \href [0]{\begingroup \@sanitize@url \@href}%
\providecommand \@href[1]{\@@startlink{#1}\@@href}%
\providecommand \@@href[1]{\endgroup#1\@@endlink}%
\providecommand \@sanitize@url [0]{\catcode `\\12\catcode `\$12\catcode
  `\&12\catcode `\#12\catcode `\^12\catcode `\_12\catcode `\%12\relax}%
\providecommand \@@startlink[1]{}%
\providecommand \@@endlink[0]{}%
\providecommand \url  [0]{\begingroup\@sanitize@url \@url }%
\providecommand \@url [1]{\endgroup\@href {#1}{\urlprefix }}%
\providecommand \urlprefix  [0]{URL }%
\providecommand \Eprint [0]{\href }%
\providecommand \doibase [0]{http://dx.doi.org/}%
\providecommand \selectlanguage [0]{\@gobble}%
\providecommand \bibinfo  [0]{\@secondoftwo}%
\providecommand \bibfield  [0]{\@secondoftwo}%
\providecommand \translation [1]{[#1]}%
\providecommand \BibitemOpen [0]{}%
\providecommand \bibitemStop [0]{}%
\providecommand \bibitemNoStop [0]{.\EOS\space}%
\providecommand \EOS [0]{\spacefactor3000\relax}%
\providecommand \BibitemShut  [1]{\csname bibitem#1\endcsname}%
\let\auto@bib@innerbib\@empty
%</preamble>
\bibitem [{\citenamefont {Weisbuch}\ \emph {et~al.}(1992)\citenamefont
  {Weisbuch}, \citenamefont {Nishioka}, \citenamefont {Ishikawa},\ and\
  \citenamefont {Arakawa}}]{Weisbuch}%
  \BibitemOpen
  \bibfield  {author} {\bibinfo {author} {\bibfnamefont {C.}~\bibnamefont
  {Weisbuch}}, \bibinfo {author} {\bibfnamefont {M.}~\bibnamefont {Nishioka}},
  \bibinfo {author} {\bibfnamefont {A.}~\bibnamefont {Ishikawa}}, \ and\
  \bibinfo {author} {\bibfnamefont {Y.}~\bibnamefont {Arakawa}},\ }\href
  {\doibase 10.1103/PhysRevLett.69.3314} {\bibfield  {journal} {\bibinfo
  {journal} {Phys. Rev. Lett.}\ }\textbf {\bibinfo {volume} {69}},\ \bibinfo
  {pages} {3314} (\bibinfo {year} {1992})}\BibitemShut {NoStop}%
\bibitem [{\citenamefont {Schneider}\ \emph {et~al.}(2013)\citenamefont
  {Schneider}, \citenamefont {Rahimi-Iman}, \citenamefont {Kim}, \citenamefont
  {Fischer}, \citenamefont {Savenko}, \citenamefont {Amthor}, \citenamefont
  {Lermer}, \citenamefont {Wolf}, \citenamefont {Worschech}, \citenamefont
  {Kulakovskii}, \citenamefont {Shelykh}, \citenamefont {Kamp}, \citenamefont
  {Reitzenstein}, \citenamefont {Forchel}, \citenamefont {Yamamoto},\ and\
  \citenamefont {Hofling}}]{Schneider:2013qr}%
  \BibitemOpen
  \bibfield  {author} {\bibinfo {author} {\bibfnamefont {C.}~\bibnamefont
  {Schneider}}, \bibinfo {author} {\bibfnamefont {A.}~\bibnamefont
  {Rahimi-Iman}}, \bibinfo {author} {\bibfnamefont {N.~Y.}\ \bibnamefont
  {Kim}}, \bibinfo {author} {\bibfnamefont {J.}~\bibnamefont {Fischer}},
  \bibinfo {author} {\bibfnamefont {I.~G.}\ \bibnamefont {Savenko}}, \bibinfo
  {author} {\bibfnamefont {M.}~\bibnamefont {Amthor}}, \bibinfo {author}
  {\bibfnamefont {M.}~\bibnamefont {Lermer}}, \bibinfo {author} {\bibfnamefont
  {A.}~\bibnamefont {Wolf}}, \bibinfo {author} {\bibfnamefont {L.}~\bibnamefont
  {Worschech}}, \bibinfo {author} {\bibfnamefont {V.~D.}\ \bibnamefont
  {Kulakovskii}}, \bibinfo {author} {\bibfnamefont {I.~A.}\ \bibnamefont
  {Shelykh}}, \bibinfo {author} {\bibfnamefont {M.}~\bibnamefont {Kamp}},
  \bibinfo {author} {\bibfnamefont {S.}~\bibnamefont {Reitzenstein}}, \bibinfo
  {author} {\bibfnamefont {A.}~\bibnamefont {Forchel}}, \bibinfo {author}
  {\bibfnamefont {Y.}~\bibnamefont {Yamamoto}}, \ and\ \bibinfo {author}
  {\bibfnamefont {S.}~\bibnamefont {Hofling}},\ }\href
  {http://dx.doi.org/10.1038/nature12036} {\bibfield  {journal} {\bibinfo
  {journal} {Nature}\ }\textbf {\bibinfo {volume} {497}},\ \bibinfo {pages}
  {348} (\bibinfo {year} {2013})}\BibitemShut {NoStop}%
\bibitem [{\citenamefont {Bhattacharya}\ \emph {et~al.}(2013)\citenamefont
  {Bhattacharya}, \citenamefont {Xiao}, \citenamefont {Das}, \citenamefont
  {Bhowmick},\ and\ \citenamefont {Heo}}]{PhysRevLett.110.206403}%
  \BibitemOpen
  \bibfield  {author} {\bibinfo {author} {\bibfnamefont {P.}~\bibnamefont
  {Bhattacharya}}, \bibinfo {author} {\bibfnamefont {B.}~\bibnamefont {Xiao}},
  \bibinfo {author} {\bibfnamefont {A.}~\bibnamefont {Das}}, \bibinfo {author}
  {\bibfnamefont {S.}~\bibnamefont {Bhowmick}}, \ and\ \bibinfo {author}
  {\bibfnamefont {J.}~\bibnamefont {Heo}},\ }\href {\doibase
  10.1103/PhysRevLett.110.206403} {\bibfield  {journal} {\bibinfo  {journal}
  {Phys. Rev. Lett.}\ }\textbf {\bibinfo {volume} {110}},\ \bibinfo {pages}
  {206403} (\bibinfo {year} {2013})}\BibitemShut {NoStop}%
\bibitem [{\citenamefont {Ballarini}\ \emph {et~al.}(2013)\citenamefont
  {Ballarini}, \citenamefont {De~Giorgi}, \citenamefont {Cancellieri},
  \citenamefont {Houdr{\'e}}, \citenamefont {Giacobino}, \citenamefont
  {Cingolani}, \citenamefont {Bramati}, \citenamefont {Gigli},\ and\
  \citenamefont {Sanvitto}}]{Ballarini:2013}%
  \BibitemOpen
  \bibfield  {author} {\bibinfo {author} {\bibfnamefont {D.}~\bibnamefont
  {Ballarini}}, \bibinfo {author} {\bibfnamefont {M.}~\bibnamefont
  {De~Giorgi}}, \bibinfo {author} {\bibfnamefont {E.}~\bibnamefont
  {Cancellieri}}, \bibinfo {author} {\bibfnamefont {R.}~\bibnamefont
  {Houdr{\'e}}}, \bibinfo {author} {\bibfnamefont {E.}~\bibnamefont
  {Giacobino}}, \bibinfo {author} {\bibfnamefont {R.}~\bibnamefont
  {Cingolani}}, \bibinfo {author} {\bibfnamefont {A.}~\bibnamefont {Bramati}},
  \bibinfo {author} {\bibfnamefont {G.}~\bibnamefont {Gigli}}, \ and\ \bibinfo
  {author} {\bibfnamefont {D.}~\bibnamefont {Sanvitto}},\ }\href
  {http://dx.doi.org/10.1038/ncomms2734} {\bibfield  {journal} {\bibinfo
  {journal} {Nature Communications}\ }\textbf {\bibinfo {volume} {4}},\
  \bibinfo {pages} {1778} (\bibinfo {year} {2013})}\BibitemShut {NoStop}%
\bibitem [{\citenamefont {Leyder}\ \emph {et~al.}(2007)\citenamefont {Leyder},
  \citenamefont {Liew}, \citenamefont {Kavokin}, \citenamefont {Shelykh},
  \citenamefont {Romanelli}, \citenamefont {Karr}, \citenamefont {Giacobino},\
  and\ \citenamefont {Bramati}}]{PhysRevLett.99.196402}%
  \BibitemOpen
  \bibfield  {author} {\bibinfo {author} {\bibfnamefont {C.}~\bibnamefont
  {Leyder}}, \bibinfo {author} {\bibfnamefont {T.~C.~H.}\ \bibnamefont {Liew}},
  \bibinfo {author} {\bibfnamefont {A.~V.}\ \bibnamefont {Kavokin}}, \bibinfo
  {author} {\bibfnamefont {I.~A.}\ \bibnamefont {Shelykh}}, \bibinfo {author}
  {\bibfnamefont {M.}~\bibnamefont {Romanelli}}, \bibinfo {author}
  {\bibfnamefont {J.~P.}\ \bibnamefont {Karr}}, \bibinfo {author}
  {\bibfnamefont {E.}~\bibnamefont {Giacobino}}, \ and\ \bibinfo {author}
  {\bibfnamefont {A.}~\bibnamefont {Bramati}},\ }\href
  {http://link.aps.org/doi/10.1103/PhysRevLett.99.196402} {\bibfield  {journal}
  {\bibinfo  {journal} {Phys. Rev. Lett.}\ }\textbf {\bibinfo {volume} {99}},\
  \bibinfo {pages} {196402} (\bibinfo {year} {2007})}\BibitemShut {NoStop}%
\bibitem [{\citenamefont {Demirchyan}\ \emph {et~al.}(2014)\citenamefont
  {Demirchyan}, \citenamefont {Chestnov}, \citenamefont {Alodjants},
  \citenamefont {Glazov},\ and\ \citenamefont {Kavokin}}]{Demirchyan}%
  \BibitemOpen
  \bibfield  {author} {\bibinfo {author} {\bibfnamefont {S.~S.}\ \bibnamefont
  {Demirchyan}}, \bibinfo {author} {\bibfnamefont {I.~Y.}\ \bibnamefont
  {Chestnov}}, \bibinfo {author} {\bibfnamefont {A.~P.}\ \bibnamefont
  {Alodjants}}, \bibinfo {author} {\bibfnamefont {M.~M.}\ \bibnamefont
  {Glazov}}, \ and\ \bibinfo {author} {\bibfnamefont {A.~V.}\ \bibnamefont
  {Kavokin}},\ }\href {\doibase 10.1103/PhysRevLett.112.196403} {\bibfield
  {journal} {\bibinfo  {journal} {Phys. Rev. Lett.}\ }\textbf {\bibinfo
  {volume} {112}},\ \bibinfo {pages} {196403} (\bibinfo {year}
  {2014})}\BibitemShut {NoStop}%
\bibitem [{\citenamefont {Liew}\ \emph {et~al.}(2013)\citenamefont {Liew},
  \citenamefont {Glazov}, \citenamefont {Kavokin}, \citenamefont {Shelykh},
  \citenamefont {Kaliteevski},\ and\ \citenamefont
  {Kavokin}}]{PhysRevLett.110.047402}%
  \BibitemOpen
  \bibfield  {author} {\bibinfo {author} {\bibfnamefont {T.~C.~H.}\
  \bibnamefont {Liew}}, \bibinfo {author} {\bibfnamefont {M.~M.}\ \bibnamefont
  {Glazov}}, \bibinfo {author} {\bibfnamefont {K.~V.}\ \bibnamefont {Kavokin}},
  \bibinfo {author} {\bibfnamefont {I.~A.}\ \bibnamefont {Shelykh}}, \bibinfo
  {author} {\bibfnamefont {M.~A.}\ \bibnamefont {Kaliteevski}}, \ and\ \bibinfo
  {author} {\bibfnamefont {A.~V.}\ \bibnamefont {Kavokin}},\ }\href
  {http://link.aps.org/doi/10.1103/PhysRevLett.110.047402} {\bibfield
  {journal} {\bibinfo  {journal} {Phys. Rev. Lett.}\ }\textbf {\bibinfo
  {volume} {110}},\ \bibinfo {pages} {047402} (\bibinfo {year}
  {2013})}\BibitemShut {NoStop}%
\bibitem [{\citenamefont {Kiselev~V.A.}(1973)}]{Razbirin}%
  \BibitemOpen
  \bibfield  {author} {\bibinfo {author} {\bibfnamefont {U.~I.}\ \bibnamefont
  {Kiselev~V.A.}, \bibfnamefont {Razbirin~B.S.}},\ }\href@noop {} {\bibfield
  {journal} {\bibinfo  {journal} {JETP Letters}\ }\textbf {\bibinfo {volume}
  {18}},\ \bibinfo {pages} {296} (\bibinfo {year} {1973})}\BibitemShut
  {NoStop}%
\bibitem [{\citenamefont {Norris}\ \emph {et~al.}(1994)\citenamefont {Norris},
  \citenamefont {Rhee}, \citenamefont {Sung}, \citenamefont {Arakawa},
  \citenamefont {Nishioka},\ and\ \citenamefont {Weisbuch}}]{Norris}%
  \BibitemOpen
  \bibfield  {author} {\bibinfo {author} {\bibfnamefont {T.~B.}\ \bibnamefont
  {Norris}}, \bibinfo {author} {\bibfnamefont {J.-K.}\ \bibnamefont {Rhee}},
  \bibinfo {author} {\bibfnamefont {C.-Y.}\ \bibnamefont {Sung}}, \bibinfo
  {author} {\bibfnamefont {Y.}~\bibnamefont {Arakawa}}, \bibinfo {author}
  {\bibfnamefont {M.}~\bibnamefont {Nishioka}}, \ and\ \bibinfo {author}
  {\bibfnamefont {C.}~\bibnamefont {Weisbuch}},\ }\href
  {http://link.aps.org/doi/10.1103/PhysRevB.50.14663} {\bibfield  {journal}
  {\bibinfo  {journal} {Phys. Rev. B}\ }\textbf {\bibinfo {volume} {50}},\
  \bibinfo {pages} {14663} (\bibinfo {year} {1994})}\BibitemShut {NoStop}%
\bibitem [{\citenamefont {Kelkar}\ \emph {et~al.}(1997)\citenamefont {Kelkar},
  \citenamefont {Kozlov}, \citenamefont {Nurmikko}, \citenamefont {Chu},
  \citenamefont {Han},\ and\ \citenamefont {Gunshor}}]{PhysRevB.56.7564}%
  \BibitemOpen
  \bibfield  {author} {\bibinfo {author} {\bibfnamefont {P.~V.}\ \bibnamefont
  {Kelkar}}, \bibinfo {author} {\bibfnamefont {V.~G.}\ \bibnamefont {Kozlov}},
  \bibinfo {author} {\bibfnamefont {A.~V.}\ \bibnamefont {Nurmikko}}, \bibinfo
  {author} {\bibfnamefont {C.-C.}\ \bibnamefont {Chu}}, \bibinfo {author}
  {\bibfnamefont {J.}~\bibnamefont {Han}}, \ and\ \bibinfo {author}
  {\bibfnamefont {R.~L.}\ \bibnamefont {Gunshor}},\ }\href {\doibase
  10.1103/PhysRevB.56.7564} {\bibfield  {journal} {\bibinfo  {journal} {Phys.
  Rev. B}\ }\textbf {\bibinfo {volume} {56}},\ \bibinfo {pages} {7564}
  (\bibinfo {year} {1997})}\BibitemShut {NoStop}%
\bibitem [{\citenamefont {Bongiovanni}\ \emph {et~al.}(1997)\citenamefont
  {Bongiovanni}, \citenamefont {Mura}, \citenamefont {Quochi}, \citenamefont
  {G\"urtler}, \citenamefont {Staehli}, \citenamefont {Tassone}, \citenamefont
  {Stanley}, \citenamefont {Oesterle},\ and\ \citenamefont
  {Houdr\'e}}]{PhysRevB.55.7084}%
  \BibitemOpen
  \bibfield  {author} {\bibinfo {author} {\bibfnamefont {G.}~\bibnamefont
  {Bongiovanni}}, \bibinfo {author} {\bibfnamefont {A.}~\bibnamefont {Mura}},
  \bibinfo {author} {\bibfnamefont {F.}~\bibnamefont {Quochi}}, \bibinfo
  {author} {\bibfnamefont {S.}~\bibnamefont {G\"urtler}}, \bibinfo {author}
  {\bibfnamefont {J.~L.}\ \bibnamefont {Staehli}}, \bibinfo {author}
  {\bibfnamefont {F.}~\bibnamefont {Tassone}}, \bibinfo {author} {\bibfnamefont
  {R.~P.}\ \bibnamefont {Stanley}}, \bibinfo {author} {\bibfnamefont
  {U.}~\bibnamefont {Oesterle}}, \ and\ \bibinfo {author} {\bibfnamefont
  {R.}~\bibnamefont {Houdr\'e}},\ }\href {\doibase 10.1103/PhysRevB.55.7084}
  {\bibfield  {journal} {\bibinfo  {journal} {Phys. Rev. B}\ }\textbf {\bibinfo
  {volume} {55}},\ \bibinfo {pages} {7084} (\bibinfo {year}
  {1997})}\BibitemShut {NoStop}%
\bibitem [{\citenamefont {Shirane}\ \emph {et~al.}(1998)\citenamefont
  {Shirane}, \citenamefont {Ramkumar}, \citenamefont {Svirko}, \citenamefont
  {Suzuura}, \citenamefont {Inouye}, \citenamefont {Shimano}, \citenamefont
  {Someya}, \citenamefont {Sakaki},\ and\ \citenamefont
  {Kuwata-Gonokami}}]{PhysRevB.58.7978}%
  \BibitemOpen
  \bibfield  {author} {\bibinfo {author} {\bibfnamefont {M.}~\bibnamefont
  {Shirane}}, \bibinfo {author} {\bibfnamefont {C.}~\bibnamefont {Ramkumar}},
  \bibinfo {author} {\bibfnamefont {Y.~P.}\ \bibnamefont {Svirko}}, \bibinfo
  {author} {\bibfnamefont {H.}~\bibnamefont {Suzuura}}, \bibinfo {author}
  {\bibfnamefont {S.}~\bibnamefont {Inouye}}, \bibinfo {author} {\bibfnamefont
  {R.}~\bibnamefont {Shimano}}, \bibinfo {author} {\bibfnamefont
  {T.}~\bibnamefont {Someya}}, \bibinfo {author} {\bibfnamefont
  {H.}~\bibnamefont {Sakaki}}, \ and\ \bibinfo {author} {\bibfnamefont
  {M.}~\bibnamefont {Kuwata-Gonokami}},\ }\href {\doibase
  10.1103/PhysRevB.58.7978} {\bibfield  {journal} {\bibinfo  {journal} {Phys.
  Rev. B}\ }\textbf {\bibinfo {volume} {58}},\ \bibinfo {pages} {7978}
  (\bibinfo {year} {1998})}\BibitemShut {NoStop}%
\bibitem [{\citenamefont {Brunetti}\ \emph {et~al.}(2006)\citenamefont
  {Brunetti}, \citenamefont {Vladimirova}, \citenamefont {Scalbert},
  \citenamefont {Nawrocki}, \citenamefont {Kavokin}, \citenamefont {Shelykh},\
  and\ \citenamefont {Bloch}}]{MV_Rabi}%
  \BibitemOpen
  \bibfield  {author} {\bibinfo {author} {\bibfnamefont {A.}~\bibnamefont
  {Brunetti}}, \bibinfo {author} {\bibfnamefont {M.}~\bibnamefont
  {Vladimirova}}, \bibinfo {author} {\bibfnamefont {D.}~\bibnamefont
  {Scalbert}}, \bibinfo {author} {\bibfnamefont {M.}~\bibnamefont {Nawrocki}},
  \bibinfo {author} {\bibfnamefont {A.~V.}\ \bibnamefont {Kavokin}}, \bibinfo
  {author} {\bibfnamefont {I.~A.}\ \bibnamefont {Shelykh}}, \ and\ \bibinfo
  {author} {\bibfnamefont {J.}~\bibnamefont {Bloch}},\ }\href
  {http://link.aps.org/doi/10.1103/PhysRevB.74.241101} {\bibfield  {journal}
  {\bibinfo  {journal} {Phys. Rev. B}\ }\textbf {\bibinfo {volume} {74}},\
  \bibinfo {pages} {241101} (\bibinfo {year} {2006})}\BibitemShut {NoStop}%
\bibitem [{\citenamefont {Liew}\ \emph {et~al.}(2014)\citenamefont {Liew},
  \citenamefont {Rubo},\ and\ \citenamefont {Kavokin}}]{PhysRevB.90.245309}%
  \BibitemOpen
  \bibfield  {author} {\bibinfo {author} {\bibfnamefont {T.~C.~H.}\
  \bibnamefont {Liew}}, \bibinfo {author} {\bibfnamefont {Y.~G.}\ \bibnamefont
  {Rubo}}, \ and\ \bibinfo {author} {\bibfnamefont {A.~V.}\ \bibnamefont
  {Kavokin}},\ }\href {\doibase 10.1103/PhysRevB.90.245309} {\bibfield
  {journal} {\bibinfo  {journal} {Phys. Rev. B}\ }\textbf {\bibinfo {volume}
  {90}},\ \bibinfo {pages} {245309} (\bibinfo {year} {2014})}\BibitemShut
  {NoStop}%
\bibitem [{\citenamefont {De~Giorgi}\ \emph {et~al.}(2014)\citenamefont
  {De~Giorgi}, \citenamefont {Ballarini}, \citenamefont {Cazzato},
  \citenamefont {Deligeorgis}, \citenamefont {Tsintzos}, \citenamefont
  {Hatzopoulos}, \citenamefont {Savvidis}, \citenamefont {Gigli}, \citenamefont
  {Laussy},\ and\ \citenamefont {Sanvitto}}]{PhysRevLett.112.113602}%
  \BibitemOpen
  \bibfield  {author} {\bibinfo {author} {\bibfnamefont {M.}~\bibnamefont
  {De~Giorgi}}, \bibinfo {author} {\bibfnamefont {D.}~\bibnamefont
  {Ballarini}}, \bibinfo {author} {\bibfnamefont {P.}~\bibnamefont {Cazzato}},
  \bibinfo {author} {\bibfnamefont {G.}~\bibnamefont {Deligeorgis}}, \bibinfo
  {author} {\bibfnamefont {S.~I.}\ \bibnamefont {Tsintzos}}, \bibinfo {author}
  {\bibfnamefont {Z.}~\bibnamefont {Hatzopoulos}}, \bibinfo {author}
  {\bibfnamefont {P.~G.}\ \bibnamefont {Savvidis}}, \bibinfo {author}
  {\bibfnamefont {G.}~\bibnamefont {Gigli}}, \bibinfo {author} {\bibfnamefont
  {F.~P.}\ \bibnamefont {Laussy}}, \ and\ \bibinfo {author} {\bibfnamefont
  {D.}~\bibnamefont {Sanvitto}},\ }\href
  {http://link.aps.org/doi/10.1103/PhysRevLett.112.113602} {\bibfield
  {journal} {\bibinfo  {journal} {Phys. Rev. Lett.}\ }\textbf {\bibinfo
  {volume} {112}},\ \bibinfo {pages} {113602} (\bibinfo {year}
  {2014})}\BibitemShut {NoStop}%
     \bibitem [{\citenamefont {PhysRevLett.113.226401}\ \emph {et~al.}(2014)\citenamefont
  {Dominici}, \citenamefont {Colas}, \citenamefont {Donati}, \citenamefont {Cuartas}, \citenamefont {Giorgi}, \citenamefont {Ballarini}, \citenamefont {Guirales}, \citenamefont {Carreno},\citenamefont {Bramati},\citenamefont {Gigli}, \citenamefont {Valle}, \citenamefont {Laussy},\ and\
  \citenamefont {Sanvitto}}]{PhysRevLett.113.226401}%
  \BibitemOpen
  \bibfield  {author} { \bibinfo {author} {\bibfnamefont {L.}~\bibnamefont {Dominici}},
  \bibinfo {author} {\bibfnamefont {D.}~\bibnamefont  {Colas}},
  \bibinfo {author} {\bibfnamefont {S.}~\bibnamefont {Donati}},
    \bibinfo {author} {\bibfnamefont {J.~P.}~\bibnamefont {Restrepo Cuartas}},
            \bibinfo {author} {\bibfnamefont {M.}~\bibnamefont {De Giorgi}},
             \bibinfo {author} {\bibfnamefont {D.}~\bibnamefont {Ballarini}},
                  \bibinfo {author} {\bibfnamefont {G.}~\bibnamefont {Guirales}},
        \bibinfo {author} {\bibfnamefont {J.~C.}~\bibnamefont {L\'opez Carre\~no}},
              \bibinfo {author} {\bibfnamefont {A.}~\bibnamefont {Bramati}},
                \bibinfo {author} {\bibfnamefont {G.}~\bibnamefont {Gigli}},
                  \bibinfo {author} {\bibfnamefont {E.}~\bibnamefont {Del Valle}},
                    \bibinfo {author} {\bibfnamefont {F.~P.}~\bibnamefont {Laussy}},                 
   \ and\
  \bibinfo {author} {\bibfnamefont {D.}~\bibnamefont {Sanvitto}},\ }\href
  {\doibase 10.1103/PhysRevLett.113.226401} {\bibfield  {journal} {\bibinfo
  {journal} {Phys. Rev. Lett.}\ }\textbf {\bibinfo {volume} {113}},\ \bibinfo
  {pages} {226401} (\bibinfo {year} {2014})}\BibitemShut {NoStop}%
     \bibitem [{\citenamefont {Colas2015}\ \emph {et~al.}(2015)\citenamefont
  {Colas}, \citenamefont {Dominici}, \citenamefont {Donati}, \citenamefont {Pervishko}, \citenamefont {Liew}, \citenamefont {Shelykh}, \citenamefont {Ballarini}, \citenamefont {Giorgi},\citenamefont {Bramati},\citenamefont {Gigli}, \citenamefont {Valle}, \citenamefont {Laussy}, \citenamefont {Kavokin},\ and\
  \citenamefont {Sanvitto}}]{Colas2015}%
  \BibitemOpen
  \bibfield  {author} {\bibinfo {author} {\bibfnamefont {D.}~\bibnamefont
  {Colas}}, \bibinfo {author} {\bibfnamefont {L.}~\bibnamefont {Dominici}},
  \bibinfo {author} {\bibfnamefont {S.}~\bibnamefont {Donati}},
    \bibinfo {author} {\bibfnamefont {A.~A.}~\bibnamefont {Pervishko}},
      \bibinfo {author} {\bibfnamefont {T.~C.~H.}~\bibnamefont {Liew}},
        \bibinfo {author} {\bibfnamefont {I.~A.}~\bibnamefont {Shelykh}},
          \bibinfo {author} {\bibfnamefont {D.}~\bibnamefont {Ballarini}},
            \bibinfo {author} {\bibfnamefont {M.}~\bibnamefont {De Giorgi}},
              \bibinfo {author} {\bibfnamefont {A.}~\bibnamefont {Bramati}},
                \bibinfo {author} {\bibfnamefont {G.}~\bibnamefont {Gigli}},
                  \bibinfo {author} {\bibfnamefont {E.}~\bibnamefont {Del Valle}},
                    \bibinfo {author} {\bibfnamefont {F.~P.}~\bibnamefont {Laussy}},
                      \bibinfo {author} {\bibfnamefont {A.~V.}~\bibnamefont {Kavokin}},                   
   \ and\
  \bibinfo {author} {\bibfnamefont {D.}~\bibnamefont {Sanvitto}},\ }\href
  {\doibase 10.1038/lsa.2015.123} {\bibfield  {journal} {\bibinfo
  {journal} {Light: Science and Applications}\ }\textbf {\bibinfo {volume} {4}},\ \bibinfo
  {pages} {e350} (\bibinfo {year} {2015})}\BibitemShut {NoStop}%
\bibitem [{\citenamefont {Takemura}\ \emph
  {et~al.}(2015{\natexlab{a}})\citenamefont {Takemura}, \citenamefont
  {Anderson}, \citenamefont {Trebaol}, \citenamefont {Biswas}, \citenamefont
  {Oberli}, \citenamefont {Portella-Oberli},\ and\ \citenamefont
  {Deveaud}}]{PhysRevB.92.235305}%
  \BibitemOpen
  \bibfield  {author} {\bibinfo {author} {\bibfnamefont {N.}~\bibnamefont
  {Takemura}}, \bibinfo {author} {\bibfnamefont {M.~D.}\ \bibnamefont
  {Anderson}}, \bibinfo {author} {\bibfnamefont {S.}~\bibnamefont {Trebaol}},
  \bibinfo {author} {\bibfnamefont {S.}~\bibnamefont {Biswas}}, \bibinfo
  {author} {\bibfnamefont {D.~Y.}\ \bibnamefont {Oberli}}, \bibinfo {author}
  {\bibfnamefont {M.~T.}\ \bibnamefont {Portella-Oberli}}, \ and\ \bibinfo
  {author} {\bibfnamefont {B.}~\bibnamefont {Deveaud}},\ }\href {\doibase
  10.1103/PhysRevB.92.235305} {\bibfield  {journal} {\bibinfo  {journal} {Phys.
  Rev. B}\ }\textbf {\bibinfo {volume} {92}},\ \bibinfo {pages} {235305}
  (\bibinfo {year} {2015}{\natexlab{a}})}\BibitemShut {NoStop}%
\bibitem [{\citenamefont {Takemura}\ \emph {et~al.}(2016)\citenamefont
  {Takemura}, \citenamefont {Anderson}, \citenamefont {Biswas}, \citenamefont
  {Navadeh-Toupchi}, \citenamefont {Oberli}, \citenamefont {Portella-Oberli},\
  and\ \citenamefont {Deveaud}}]{PhysRevB.94.195301}%
  \BibitemOpen
  \bibfield  {author} {\bibinfo {author} {\bibfnamefont {N.}~\bibnamefont
  {Takemura}}, \bibinfo {author} {\bibfnamefont {M.~D.}\ \bibnamefont
  {Anderson}}, \bibinfo {author} {\bibfnamefont {S.}~\bibnamefont {Biswas}},
  \bibinfo {author} {\bibfnamefont {M.}~\bibnamefont {Navadeh-Toupchi}},
  \bibinfo {author} {\bibfnamefont {D.~Y.}\ \bibnamefont {Oberli}}, \bibinfo
  {author} {\bibfnamefont {M.~T.}\ \bibnamefont {Portella-Oberli}}, \ and\
  \bibinfo {author} {\bibfnamefont {B.}~\bibnamefont {Deveaud}},\ }\href
  {\doibase 10.1103/PhysRevB.94.195301} {\bibfield  {journal} {\bibinfo
  {journal} {Phys. Rev. B}\ }\textbf {\bibinfo {volume} {94}},\ \bibinfo
  {pages} {195301} (\bibinfo {year} {2016})}\BibitemShut {NoStop}%
\bibitem [{\citenamefont {Takemura}\ \emph
  {et~al.}(2015{\natexlab{b}})\citenamefont {Takemura}, \citenamefont
  {Trebaol}, \citenamefont {Anderson}, \citenamefont {Kohnle}, \citenamefont
  {L\'eger}, \citenamefont {Oberli}, \citenamefont {Portella-Oberli},\ and\
  \citenamefont {Deveaud}}]{PhysRevB.92.125415}%
  \BibitemOpen
  \bibfield  {author} {\bibinfo {author} {\bibfnamefont {N.}~\bibnamefont
  {Takemura}}, \bibinfo {author} {\bibfnamefont {S.}~\bibnamefont {Trebaol}},
  \bibinfo {author} {\bibfnamefont {M.~D.}\ \bibnamefont {Anderson}}, \bibinfo
  {author} {\bibfnamefont {V.}~\bibnamefont {Kohnle}}, \bibinfo {author}
  {\bibfnamefont {Y.}~\bibnamefont {L\'eger}}, \bibinfo {author} {\bibfnamefont
  {D.~Y.}\ \bibnamefont {Oberli}}, \bibinfo {author} {\bibfnamefont {M.~T.}\
  \bibnamefont {Portella-Oberli}}, \ and\ \bibinfo {author} {\bibfnamefont
  {B.}~\bibnamefont {Deveaud}},\ }\href {\doibase 10.1103/PhysRevB.92.125415}
  {\bibfield  {journal} {\bibinfo  {journal} {Phys. Rev. B}\ }\textbf {\bibinfo
  {volume} {92}},\ \bibinfo {pages} {125415} (\bibinfo {year}
  {2015}{\natexlab{b}})}\BibitemShut {NoStop}%
\bibitem [{\citenamefont {Shah}(1999)}]{Shah1999}%
  \BibitemOpen
  \bibfield  {author} {\bibinfo {author} {\bibfnamefont {J.}~\bibnamefont
  {Shah}},\ }\href@noop {} {\emph {\bibinfo {title} {Ultrafast Spectroscopy of
  Semiconductors and Semiconductor Nanostructures, 2nd ed.}}}\ (\bibinfo
  {publisher} {Springer, Berlin},\ \bibinfo {year} {1999})\BibitemShut
  {NoStop}%
\bibitem [{\citenamefont {Ivchenko}(2005)}]{ivchenko05a}%
  \BibitemOpen
  \bibfield  {author} {\bibinfo {author} {\bibfnamefont {E.~L.}\ \bibnamefont
  {Ivchenko}},\ }\href@noop {} {\emph {\bibinfo {title} {Optical spectroscopy
  of semiconductor nanostructures}}}\ (\bibinfo  {publisher} {Alpha Science,
  Harrow UK},\ \bibinfo {year} {2005})\BibitemShut {NoStop}%
\bibitem [{\citenamefont {Fu}\ \emph {et~al.}(1997)\citenamefont {Fu},
  \citenamefont {Willander}, \citenamefont {Ivchenko},\ and\ \citenamefont
  {Kiselev}}]{PhysRevB.55.9872}%
  \BibitemOpen
  \bibfield  {author} {\bibinfo {author} {\bibfnamefont {Y.}~\bibnamefont
  {Fu}}, \bibinfo {author} {\bibfnamefont {M.}~\bibnamefont {Willander}},
  \bibinfo {author} {\bibfnamefont {E.~L.}\ \bibnamefont {Ivchenko}}, \ and\
  \bibinfo {author} {\bibfnamefont {A.~A.}\ \bibnamefont {Kiselev}},\ }\href
  {\doibase 10.1103/PhysRevB.55.9872} {\bibfield  {journal} {\bibinfo
  {journal} {Phys. Rev. B}\ }\textbf {\bibinfo {volume} {55}},\ \bibinfo
  {pages} {9872} (\bibinfo {year} {1997})}\BibitemShut {NoStop}%
\bibitem [{\citenamefont {Rochat}\ \emph {et~al.}(2000)\citenamefont {Rochat},
  \citenamefont {Ciuti}, \citenamefont {Savona}, \citenamefont {Piermarocchi},
  \citenamefont {Quattropani},\ and\ \citenamefont
  {Schwendimann}}]{PhysRevB.61.13856}%
  \BibitemOpen
  \bibfield  {author} {\bibinfo {author} {\bibfnamefont {G.}~\bibnamefont
  {Rochat}}, \bibinfo {author} {\bibfnamefont {C.}~\bibnamefont {Ciuti}},
  \bibinfo {author} {\bibfnamefont {V.}~\bibnamefont {Savona}}, \bibinfo
  {author} {\bibfnamefont {C.}~\bibnamefont {Piermarocchi}}, \bibinfo {author}
  {\bibfnamefont {A.}~\bibnamefont {Quattropani}}, \ and\ \bibinfo {author}
  {\bibfnamefont {P.}~\bibnamefont {Schwendimann}},\ }\href {\doibase
  10.1103/PhysRevB.61.13856} {\bibfield  {journal} {\bibinfo  {journal} {Phys.
  Rev. B}\ }\textbf {\bibinfo {volume} {61}},\ \bibinfo {pages} {13856}
  (\bibinfo {year} {2000})}\BibitemShut {NoStop}%
\bibitem [{\citenamefont {Carmichael}(1993)}]{carmichael_book}%
  \BibitemOpen
  \bibfield  {author} {\bibinfo {author} {\bibfnamefont {H.}~\bibnamefont
  {Carmichael}},\ }\href@noop {} {\emph {\bibinfo {title} {An open systems
  approach to quantum optics}}}\ (\bibinfo  {publisher} {Springer-Verlag
  Berlin},\ \bibinfo {year} {1993})\BibitemShut {NoStop}%
\bibitem [{Note1()}]{Note1}%
  \BibitemOpen
  \bibinfo {note} {We note that to describe the evolution of the excitonic mode
  the so-called excitonic Bloch equations were used in recent works~\cite
  {PhysRevB.61.13856, PhysRevB.92.235305}. In this approach there is a third
  equation, which accounts for additional decay channels of exciton population
  $N_x$ as compared to polarization decay $\Gamma $. The inclusion of these
  addition channels does not change general conclusions of our work, therefore
  we neglect them and use $N_x = |P|^2$.}\BibitemShut {Stop}%
\bibitem [{\citenamefont {Schmitt-Rink}\ \emph {et~al.}(1991)\citenamefont
  {Schmitt-Rink}, \citenamefont {Mukamel}, \citenamefont {Leo}, \citenamefont
  {Shah},\ and\ \citenamefont {Chemla}}]{PhysRevA.44.2124}%
  \BibitemOpen
  \bibfield  {author} {\bibinfo {author} {\bibfnamefont {S.}~\bibnamefont
  {Schmitt-Rink}}, \bibinfo {author} {\bibfnamefont {S.}~\bibnamefont
  {Mukamel}}, \bibinfo {author} {\bibfnamefont {K.}~\bibnamefont {Leo}},
  \bibinfo {author} {\bibfnamefont {J.}~\bibnamefont {Shah}}, \ and\ \bibinfo
  {author} {\bibfnamefont {D.~S.}\ \bibnamefont {Chemla}},\ }\href {\doibase
  10.1103/PhysRevA.44.2124} {\bibfield  {journal} {\bibinfo  {journal} {Phys.
  Rev. A}\ }\textbf {\bibinfo {volume} {44}},\ \bibinfo {pages} {2124}
  (\bibinfo {year} {1991})}\BibitemShut {NoStop}%
\bibitem [{\citenamefont {Hopfield}(1958)}]{PhysRev.112.1555}%
  \BibitemOpen
  \bibfield  {author} {\bibinfo {author} {\bibfnamefont {J.~J.}\ \bibnamefont
  {Hopfield}},\ }\href {\doibase 10.1103/PhysRev.112.1555} {\bibfield
  {journal} {\bibinfo  {journal} {Phys. Rev.}\ }\textbf {\bibinfo {volume}
  {112}},\ \bibinfo {pages} {1555} (\bibinfo {year} {1958})}\BibitemShut
  {NoStop}%
\bibitem [{\citenamefont {Trifonov}\ \emph
  {et~al.}(2015{\natexlab{a}})\citenamefont {Trifonov}, \citenamefont
  {Korotan}, \citenamefont {Kurdyubov}, \citenamefont {Gerlovin}, \citenamefont
  {Ignatiev}, \citenamefont {Efimov}, \citenamefont {Eliseev}, \citenamefont
  {Petrov}, \citenamefont {Dolgikh}, \citenamefont {Ovsyankin},\ and\
  \citenamefont {Kavokin}}]{PhysRevB.91.115307}%
  \BibitemOpen
  \bibfield  {author} {\bibinfo {author} {\bibfnamefont {A.~V.}\ \bibnamefont
  {Trifonov}}, \bibinfo {author} {\bibfnamefont {S.~N.}\ \bibnamefont
  {Korotan}}, \bibinfo {author} {\bibfnamefont {A.~S.}\ \bibnamefont
  {Kurdyubov}}, \bibinfo {author} {\bibfnamefont {I.~Y.}\ \bibnamefont
  {Gerlovin}}, \bibinfo {author} {\bibfnamefont {I.~V.}\ \bibnamefont
  {Ignatiev}}, \bibinfo {author} {\bibfnamefont {Y.~P.}\ \bibnamefont
  {Efimov}}, \bibinfo {author} {\bibfnamefont {S.~A.}\ \bibnamefont {Eliseev}},
  \bibinfo {author} {\bibfnamefont {V.~V.}\ \bibnamefont {Petrov}}, \bibinfo
  {author} {\bibfnamefont {Y.~K.}\ \bibnamefont {Dolgikh}}, \bibinfo {author}
  {\bibfnamefont {V.~V.}\ \bibnamefont {Ovsyankin}}, \ and\ \bibinfo {author}
  {\bibfnamefont {A.~V.}\ \bibnamefont {Kavokin}},\ }\href {\doibase
  10.1103/PhysRevB.91.115307} {\bibfield  {journal} {\bibinfo  {journal} {Phys.
  Rev. B}\ }\textbf {\bibinfo {volume} {91}},\ \bibinfo {pages} {115307}
  (\bibinfo {year} {2015}{\natexlab{a}})}\BibitemShut {NoStop}%
\bibitem [{\citenamefont {Trifonov}\ \emph
  {et~al.}(2015{\natexlab{b}})\citenamefont {Trifonov}, \citenamefont
  {Gerlovin}, \citenamefont {Ignatiev}, \citenamefont {Yugova}, \citenamefont
  {Cherbunin}, \citenamefont {Efimov}, \citenamefont {Eliseev}, \citenamefont
  {Petrov}, \citenamefont {Lovtcius},\ and\ \citenamefont
  {Kavokin}}]{PhysRevB.92.201301}%
  \BibitemOpen
  \bibfield  {author} {\bibinfo {author} {\bibfnamefont {A.~V.}\ \bibnamefont
  {Trifonov}}, \bibinfo {author} {\bibfnamefont {I.~Y.}\ \bibnamefont
  {Gerlovin}}, \bibinfo {author} {\bibfnamefont {I.~V.}\ \bibnamefont
  {Ignatiev}}, \bibinfo {author} {\bibfnamefont {I.~A.}\ \bibnamefont
  {Yugova}}, \bibinfo {author} {\bibfnamefont {R.~V.}\ \bibnamefont
  {Cherbunin}}, \bibinfo {author} {\bibfnamefont {Y.~P.}\ \bibnamefont
  {Efimov}}, \bibinfo {author} {\bibfnamefont {S.~A.}\ \bibnamefont {Eliseev}},
  \bibinfo {author} {\bibfnamefont {V.~V.}\ \bibnamefont {Petrov}}, \bibinfo
  {author} {\bibfnamefont {V.~A.}\ \bibnamefont {Lovtcius}}, \ and\ \bibinfo
  {author} {\bibfnamefont {A.~V.}\ \bibnamefont {Kavokin}},\ }\href {\doibase
  10.1103/PhysRevB.92.201301} {\bibfield  {journal} {\bibinfo  {journal} {Phys.
  Rev. B}\ }\textbf {\bibinfo {volume} {92}},\ \bibinfo {pages} {201301}
  (\bibinfo {year} {2015}{\natexlab{b}})}\BibitemShut {NoStop}%
\bibitem [{\citenamefont {Gilliot}\ \emph {et~al.}(1999)\citenamefont
  {Gilliot}, \citenamefont {Brinkmann}, \citenamefont {Kudrna}, \citenamefont
  {Cr\'egut}, \citenamefont {L\'evy}, \citenamefont {Arnoult}, \citenamefont
  {Cibert},\ and\ \citenamefont {Tatarenko}}]{PhysRevB.60.5797}%
  \BibitemOpen
  \bibfield  {author} {\bibinfo {author} {\bibfnamefont {P.}~\bibnamefont
  {Gilliot}}, \bibinfo {author} {\bibfnamefont {D.}~\bibnamefont {Brinkmann}},
  \bibinfo {author} {\bibfnamefont {J.}~\bibnamefont {Kudrna}}, \bibinfo
  {author} {\bibfnamefont {O.}~\bibnamefont {Cr\'egut}}, \bibinfo {author}
  {\bibfnamefont {R.}~\bibnamefont {L\'evy}}, \bibinfo {author} {\bibfnamefont
  {A.}~\bibnamefont {Arnoult}}, \bibinfo {author} {\bibfnamefont
  {J.}~\bibnamefont {Cibert}}, \ and\ \bibinfo {author} {\bibfnamefont
  {S.}~\bibnamefont {Tatarenko}},\ }\href {\doibase 10.1103/PhysRevB.60.5797}
  {\bibfield  {journal} {\bibinfo  {journal} {Phys. Rev. B}\ }\textbf {\bibinfo
  {volume} {60}},\ \bibinfo {pages} {5797} (\bibinfo {year}
  {1999})}\BibitemShut {NoStop}%
\bibitem [{\citenamefont {Mitsunaga}\ and\ \citenamefont
  {Tang}(1987)}]{PhysRevA.35.1720}%
  \BibitemOpen
  \bibfield  {author} {\bibinfo {author} {\bibfnamefont {M.}~\bibnamefont
  {Mitsunaga}}\ and\ \bibinfo {author} {\bibfnamefont {C.~L.}\ \bibnamefont
  {Tang}},\ }\href {\doibase 10.1103/PhysRevA.35.1720} {\bibfield  {journal}
  {\bibinfo  {journal} {Phys. Rev. A}\ }\textbf {\bibinfo {volume} {35}},\
  \bibinfo {pages} {1720} (\bibinfo {year} {1987})}\BibitemShut {NoStop}%
\end{thebibliography}


\end{document}